\documentstyle[12pt]{article}
 \let\section=\subsection
 \let\subsection=\subsubsection
 \def\subsubsection#1{\subsection{#1}
   \par\note{CAUTION: subsection=subsubsection !}\par}
\newlength{\lll}
\newlength{\lla}
\newcommand{\preprint}[2]{\begin{table}[t]
            \begin{flushright} #1 \end{flushright}\vspace{#2}
            \end{table}}%

\preprint{DESY 92-010 \\ January 1992}{-12mm}
\title{\vbox{\vspace{-7mm}}
High Precision Verification of the Kosterlitz Thouless Scenario%
\footnote{Talk presented by K.\ Pinn
at the International Symposium for
Lattice Field Theory, Tsukuba, Japan, November 1991}}
\author{\vbox{\vspace{0mm}}
   {\bf Martin Hasenbusch$^{1}$,
        Mihail Marcu$^{2,3}$} and
   {\bf Klaus Pinn$^{4}$ }\\[6mm]
$^1\,$Fachbereich Physik, Abt.\ Theoretische Physik,
      Universit\"at Kaiserslautern,\\[-1mm]
      Postfach 3049, D-6750 Kaiserslautern, Germany\\[4mm]
$^2\,$II. Institut f\"ur Theoretische Physik, Universit\"at Hamburg,
      \\[-1mm]
      Luruper Chaussee 149, D-2000 Hamburg 50, Germany\\[4mm]
$^3\,$School of Physics and Astronomy,\\[-1mm]
      Raymond and Beverly Sackler Faculty of Exact Sciences,\\[-1mm]
      Tel Aviv University, 69978 Tel Aviv, Israel\\[4mm]
$^4\,$Institut f\"ur Theoretische Physik I, Universit\"at M\"unster,\\
      Wilhelm-Klemm-Str.\ 9, D-4400 M\"unster, Germany}
\date{}
\begin{document}
\maketitle \hfill
%\vspace{2cm}
%
\begin{abstract} \normalsize
We verify the Kosterlitz Thouless scenario for three different
SOS (solid-on-solid) models, including the dual transforms of
XY-models with Villain and with cosine action.
The method is based on a matching of the
renormalization group (RG) flow of the candidate models with the flow
of a bona fide KT model, the exactly solvable BCSOS model.
We obtain high precision estimates
for the critical couplings and other non-universal
quantities.
\end{abstract}  \hfill
\thispagestyle{empty}
%------------------------------------------------------------------------
\newpage
\noindent
For a large class of two-dimensional statistical models,
the unambigious verification
of the Kosterlitz Thouless (KT) scenario is still an open problem.
The most recent large scale Monte Carlo studies of the XY model
\cite{janke,guptanew}
clearly favor a KT against a second order transition.
However, systematic errors are not yet under control.
Here we present an alternative approach. Our method
exploits the fact that the BCSOS (body centered solid-on-solid)
model can be solved exactly
and has been {\em proven} to exhibit a KT transition \cite{exact}.
The idea is to verify the KT scenario for an SOS model
by demonstrating that its asymptotic RG flow at criticality
matches with the flow of the critical BCSOS model.
%========================================================================

\vspace{2mm}
\noindent {\bf The models.}
References for reviews on SOS models can be found in \cite{ouralg}.
The models to be defined below live on two-dimensional
square lattices with periodic boundary conditions.
The discrete Gaussian model (DG) is dual to the XY model with
Villain action. The spins $h_i$ take integer values.
The partition function is
$$
Z = \sum_{h} \exp(-K^{DG} \sum_{<i,j>} (h_i - h_j)^2) \, ,
$$
where $i$ and $j$ are nearest neighbor points.
The dual of the XY model with cosine action
also has integer valued spins $h_i$, and partition function
$$
Z = \sum_{h} \prod_{<i,j>} I_{|h_i - h_j|}(\beta^{XY}) \, ,
$$
where the $I_n$ are modified Bessel functions.
The ASOS model is defined by the partion function
$$
Z = \sum_{h} \exp(-K^{ASOS} \sum_{<i,j>} |h_i - h_j|) \, .
$$
The BCSOS model was introduced by van
Beijeren \cite{beijeren77}. The lattice
is divided in even and odd sites, like a checker board.
Spins on odd sites take values of the form $2n+1/2$, spins
on even sites are of the form $2n-1/2$, $n$ integer.
The partition function can be written as
$$
Z = \sum_{h} \exp(-K^{BCSOS} \sum_{[i,k]} |h_i - h_k|) \, ,
$$
where $i$ and $k$ are next to nearest (i.e.\ diagonal) neighbors.
Nearest neighbor spins $h_i$ and $h_j$ obey the constraint
$|h_i - h_j| = 1$.
The critical (roughening) coupling is
$K_R^{BCSOS} = \frac12 \ln2$.
The BCSOS model is equivalent to a special case of the six vertex model.
The critical behavior of several quantities is exactly known
and follows the predictions of KT theory \cite{exact}.
%=======================================================================

\vspace{2mm}
\noindent {\bf Matching.}
Universality, first introduced as the coincidence of the
critical indices, can be understood as a convergence
of the RG flow
to a universal (model independent) flow as
$K \rightarrow K_c$ and the number of block spin transformations
goes to infinity. The RG flow can be monitored
by studying correlations functions of block spins
$\phi_i = B^{-2} \sum_{j \in i} h_j$ with increasing block size $B$.

We simulated the BCSOS model at the roughening coupling
$K_R^{BCSOS}$ on $L \times L$ lattices,
with $L=12,16,24,32,48,64,96,128$.
We measured block spin functions on block systems
of size $l \times l$, with $l=1,2,4$ ($L = l \, B$).
The statistics was typically a few million single cluster updates
(see the brief algorithm discussion at the end).

Motivated by KT theory, we chose
as a monitor for the flow of the kinetic term the quantities
$%
A_n = \langle (\phi_i-\phi_j)^2 \rangle, \; n=1,2,
$
where $i$ and $j$ are nearest neighbors on the block lattice for
$n=1$, and next to nearest neighbors for $n=2$.
As a monitor for the ``fugacity''
(periodic perturbation of a massless free field)
we chose
$
A_{n+2} = \langle \cos ( 2\pi n \phi_i) \rangle, \; n=1,2,3 \, .
$

There are two parameters to be adjusted in order
to match the RG flow of one of the SOS models with that of the
critical BCSOS model. First, matching can only occur if
$K^{SOS}=K^{SOS}_R$ (the roughening coupling of the SOS model).
Secondly, one has to adjust
the ratio of the block sizes $b_m^{SOS}=B^{SOS}/B^{BCSOS}$. The
freedom to set $b_m^{SOS} \neq 1$ is necessary in order to compensate
for the different positions of the ``bare'' actions in the
Kosterlitz-Thouless flow diagram.
Matching occurs if the following condition holds:
There exists a $b_m^{SOS}$ and a $K^{BCSOS}_R$ such that for
all $i$ and for all $l$
$$
%\begin{array}{l}
A_{i,l}^{SOS}(b_m^{SOS} B^{BCSOS},K^{SOS}_R) \rightarrow
  A_{i,l}^{BCSOS}(B^{BCSOS},K^{BCSOS}_R)
%\end{array}
$$
in the limit of large $B^{BCSOS}$ (moderate $B's$ are
sufficient in practice). Here
$A_{i,l}^{SOS}(B,K)$ denotes the observable $A_i$
evaluated in the SOS model at coupling $K$ on a block system
of size $l \times l$, each block being of size $B \times B$.
Notice that no additional wave function renormalization factors
are expected, since the minima of the effective potential of the
blocked system are fixed. This is due to the fact that, at all blocking
levels, the models maintain the global symmetry of shifting all spins by
an integer.

Finite size effects are exactly cancelled since the
block systems to be compared always have the same number of blocks.
Furthermore, since the blocks themselves are already large, we expect
the matching values of $K_R^{SOS}$ and $b_m^{SOS}$ to stabilize
for small $l$ already.

In order to determine $K_R^{SOS}$ we considered,
for fixed $L^{SOS}$ and $l$, the following two equations:
$$
%\begin{array}{l}\label{eqq}
A_{i,l}^{SOS}(B^{SOS},K^{SOS}_i) =
 A_{i,l}^{BCSOS}(B^{BCSOS},K^{BCSOS}_R) \, , \quad i=1,3 \; .
%\end{array}
$$
For each of the available values of $B^{BCSOS}$ we solved
these equations numerically for $K^{SOS}_i$.
The $A_{i,l}^{SOS}$ could be computed for a whole range of couplings
with the help of the Swendsen-Ferrenberg method \cite{swefer}.
Thus we got two values $K^{SOS}_1$ and
$K^{SOS}_3$, which were in general not identical
(matching occurs only for a specific choice of $b_m^{SOS}$).
In a second step we plotted the values of $K^{SOS}_1$ and
$K^{SOS}_3$ as function of $B^{BCSOS}$. To obtain continuous curves,
we interpolated linearly in $\log B^{BCSOS}$.
The intersection of the two curves
$K^{SOS}_1(\log B^{BCSOS})$ and
$K^{SOS}_3(\log B^{BCSOS})$
then uniquely determined an estimate for
the roughening coupling $K^{SOS}_R$, and,
in addition, for $b_m^{SOS}$. This completes the matching of the
SOS and BCSOS flows at the roughening transition for given
$L^{SOS}$ and $l$.
As an example we show in table \ref{tab1} the results
for the dual of the XY model with cosine action.

%-----------------------------------------------------------------------
\begin{table}
\caption{ $\beta_R$ and $b_m=B^{XY}/B^{BCSOS}$ for the dual of the
  XY model with cosine action as obtained from the matching of
  $A_1$ and $A_3$}
  \label{tab1}
   \begin{center}
   \begin{tabular}{|c|c|c|c|l|}
   \hline
   $L^{XY}$ & $l$ & $\beta_R$ & $L^{BCSOS}$ &  $b_m$ \\
   \hline
 16 & 2 &1.1220(12) & 19.0(1.1) &   0.84(5) \\
 16 & 4 &1.1257(8) & 21.3(4)& 0.75(1)  \\
 \hline
 24 & 2 &1.1214(13) &  26.5(2.3) &  0.91(7) \\
 24 & 4 &1.1225(8) & 28.6(8)& 0.84(2)  \\
 \hline
 32 & 2 &1.1211(12) &  34.5(4.3) &  0.93(10) \\
 32 & 4 &1.1214(8) & 37.4(1.5) & 0.85(3)  \\
 \hline
 48 & 2 &1.1199(11) &  54.1(9.6) &  0.89(13)  \\
 48 & 4 &1.1205(7) & 53.6(2.1) & 0.89(3)  \\
 \hline
 64 & 2 &1.1212(11) &  78.(14.) &  0.82(12) \\
 64 & 4 &1.1201(7) & 72.1(3.9) & 0.89(5)  \\
 \hline
 96 & 2 &1.1189(11) & 108.(17.) & 0.89(12) \\
 96 & 4 &1.1194(7) & 100.9(8.0) & 0.95(7) \\
 \hline
   \end{tabular}
  \end{center}
 \end{table}
%-----------------------------------------------------------------------

For all the three SOS models considered,
the results for the roughening coupling $K_R$ obtained for the various
lattice sizes $L$ and sizes $l$ of the blocked system are consistent
with one another within statistical errors.
Only the couplings for $l=4$
on the smallest two lattices sizes and for $l=2$ on the smallest
lattices slightly deviate from the rest.
The same is true for the $b_m$.
This indicates an extremely fast
convergence to a universal RG flow of the models.
To estimate $K_R$ for the three
models we averaged the values obtained from the largest $L^{SOS}$
we considered both for $l=2$ and $l=4$, and from the second largest
$L^{SOS}$ for $l=2$ only. We arrive at the following results:
$$
\begin{array}{llllll} \label{couplings}
 \beta_R^{XY} &=& 1.1197(5), \quad \quad & b_m^{XY} &=& 0.89(5) \\
 K_R^{DG}  &=& 0.6645(6),     & b_m^{DG} &=& 0.31(2)  \\
 K_R^{ASOS} &=& 0.8061(3),    & b_m^{ASOS} &=& 2.8(3)
\end{array}
$$
The errors are statistical errors.
Systematic errors due to
deviations from the universal flow should be much smaller.

In order to check the universality of the matching we evaluated
the observables $A_2$, $A_4$ and $A_5$ at the critical couplings
$K_R$ determined above from $A_1$ and $A_3$ alone.
The results show that (within errors) matching occurs
for all observables.
%=======================================================================

\vspace{2mm}
\noindent {\bf Other non-universal constants.}
The method allows to determine other non-universal constants
appearing in the formulas for the divergence of observables
near the roughening transition \cite{bigmatch}.
For the constants in the asymptotic formula
%#%#
$\xi = A \exp\left(C \vert \frac{ K-K_R }{K_R}\vert^{1/2}\right)$
we find:
$$
\begin{array}{llllll}
A^{XY} &=& 0.223(13), \quad \quad  & C^{XY} &=& 1.78(2)    \\
A^{DG} &=& 0.078(5),   & C^{DG} &=& 2.44(3)    \\
A^{ASOS} &=& 0.70(8),  & C^{ASOS} &=& 1.14(2)
\end{array}
$$
(for the XY model we used
  $\vert \frac{\beta-\beta_R}{\beta_R}\vert$  in place of
  $\vert \frac{ K-K_R }{K_R}\vert$\,).

Our results compare quite well with results from
other Monte Carlo simulations
\cite{janke,guptanew} if the systematic
errors in these studies are taken into account.
Actually, our statistical errors are considerably smaller.
%========================================================================

\vspace{2mm}
\noindent{\bf Algorithms.}
For the algorithms used in the simulation of
the DG, the dual XY and the ASOS models we used algorithms as described
in \cite{ouralg}.
Details of the BCSOS algorithm, which is also based on the ideas
developed in \cite{ouralg}, will be described elsewhere \cite{vmrbcsos}.
Our BCSOS algorithm has a critical dynamical exponent $\approx 1$.
Therefore more than two thirds of the CPU time we used for this study
were spent for the BCSOS simulations.
With a recently developed new algorithm for the six vertex model,
the necessary resources would have been considerably
smaller \cite{hgebcsos}.
%========================================================================

The matching method was also applied to the Ising interface, for
which also a cluster algorithm was developed \cite{isiclust}.
Results on this work will be reported elsewhere \cite{isimatch}.
%========================================================================

\vspace{2mm}
\noindent{\bf Acknowledgements.}
This work was supported in part
by the Deutsche For\-schungs\-ge\-mein\-schaft (DFG),
by the German-Israeli Foundation for Research and Development (GIF),
and by the Basic Research Foundation of The Israel Academy
of Sciences and Humanities.
The numerical simulations were performed at the HLRZ in J\"ulich
and at the Regionales Hochschulrechenzentrum Kaiserslautern (RHRK).
%========================================================================

%========================================================================
\end{document}